\documentclass[12pt,a4paper]{article}
\usepackage[dvips]{graphicx}
\usepackage{amsmath,amsthm,amsfonts,cite,amssymb}
\usepackage[T2A]{fontenc}
\usepackage[cp1251]{inputenc}
\usepackage[english]{babel}

\def \bla{\boldsymbol\lambda}
\def \bpi{\boldsymbol\pi}
\def \bmu{\boldsymbol\mu}
\def \la{\lambda}
\def \CV{\mathcal V}
\def \BZ{\mathbb{Z}}
\def \BN{\mathbb{N}}

\begin{document}
\title{
$${}$$\\
{\bf Combinatorial Interpretation of the Scalar Products of State Vectors of  Integrable Models}}
\author{{\bf N.~M.~Bogoliubov,
C.~Malyshev}\\[0.5cm]
{\it\small St.-Petersburg Department of V.~A.~Steklov Mathematical Institute RAS}\\
{\it\small Fontanka 27, St.-Petersburg,
191023, Russia} }

\date{}

\maketitle
\hrule
\vskip0.5cm
{\small \begin{abstract} \noindent
The representation of the Bethe wave functions of certain integrable models via the Schur functions allows to apply the well-developed theory of the symmetric functions to the calculation of the thermal correlation functions. The algebraic relations arising in the calculation of the scalar products and the correlation functions are based on the Binet-Cauchy formula for the Schur functions. We provide a combinatorial interpretation of the formula for the scalar products of the Bethe state-vectors in terms of nests of the self-avoiding lattice paths constituting the so-called watermelon configurations. The interpretation proposed is, in its turn, related to the enumeration of the boxed plane partitions. \end{abstract}}

\vskip0.5cm
\hrule

\vskip1.5cm

\section{Introduction}

The symmetric functions, the Young diagrams, the boxed plane partitions, and the vicious walkers \cite{macd, ful, bres, forr1} play an important role in the contemporary theoretical physics \cite{zinn, ok, hik, resh, forr}.
The $N$-particle wave functions of a certain class of integrable  models on a chain are expressed in terms of Schur functions \cite{bogol, bogoltim, b1, b2, b4, b5, b6, prep}. The Schur functions are defined by the Jacobi-Trudi relation:
\begin{equation}
S_{\bla} ({\textbf x})\,\equiv\,
\displaystyle{ S_{\bla} (x_1, x_2, \dots,
x_N) \,
\equiv\,\frac{\det(x_j^{\la_k+N-k})_{1\leq
j,k\leq N}}{\CV_N({\textbf x})}} \,,
\label{sch}
\end{equation}
where $\CV_N({\textbf x})$ is the Vandermonde determinant
\begin{equation} \CV_N({\textbf x})\,\equiv\,
\det(x_j^{N-k})_{1\leq j, k\leq N}\,=\,
\prod_{1\leq m<l\leq N}(x_l-x_m)\,.
\label{spxx1}
\end{equation}
A partition $\bla \equiv (\la_1, \la_2, \ldots \la_N)$ is a nonincreasing sequence of nonnegative integers, $M \geq \la_1 \geq \la_2 \geq \ldots \geq \la_N \geq 0$, called the parts of $\bla$. Partition $\bla$ can be represented by Young diagram as an arrangement of squares with the coordinates $(i,j)$ so that $1\leq j \leq \la_i$ \cite{macd}.

For the bosonic models defined on a chain of $M+1$ sites there is one-to-one correspondence between a set of occupation numbers $\{n_M, n_{M-1} \ldots , n_1, n_0\}$ and the partition $\bla=(M^{n_M}, (M-1)^{n_{M-1}}, \ldots, 1^{n_1}, 0^{n_0})$, where each number $S$ appears $n_S$ times in ${\bla}$. For the Heisenberg spin-$\frac 12$ chains of $M+N$ sites the coordinates of the spin ``down'' states (``particles'')  constitute a strict decreasing partition ${\bmu} = (\mu_1, \mu_2,\,\dots\,,
\mu_N)$, where $M+N-1\geq \mu_1 > \mu_2 >\,\dots\,>
\mu_N \geq 0$. The parts of ${\bmu}$ and ${\bla}$ are related: $\mu_j=\la_j+N-j$.

The bijection between the coordinates of the particles and the Young diagram of the partition $\bla$ is demonstrated in Fig.~1.
\begin{figure}[h]
\center
\includegraphics {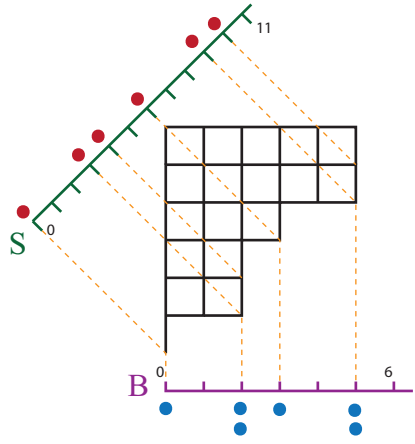}
\caption{Configurations of spins and bosons on chains and the corresponding Young
diagram of the related partition $\bla=(5,5,3,2,2,0)$.}
\end{figure}

Calculation of the correlation functions of integrable models of special type, \cite{bogol, bogoltim, b1, b2, b5, b6, prep}, is based on the Binet-Cauchy formula:
\begin{equation}
\sum_{\bla \subseteq
{{M}^N}} S_{\bla}
(\textbf{x}) S_{\bla} (\textbf{y})\,=\,\frac{\det (M_{k j})_{1\le k, j\le N} }{\CV_N(\textbf{x})\CV_N(\textbf{y})} \,,
  \label{schr}
\end{equation}
where the summation is over all partitions $\bla$ with at most $N$ parts,  each of which is less than or equal to $M$, and ${\CV}_N(\textbf{x})$ is the Vandermonde determinant (\ref{spxx1}). The entries $M_{k j}$ in (\ref{schr}) are:
\begin{equation}\label{tt}
M_{k j}=\frac{1-(x_k y_j)^{M+N}}{1-x_k y_j}\,.
\end{equation}
The calculation of the scalar products of $N$-particle Bethe state-vectors \cite{b5, b6, prep} is based on Eq.~(\ref{schr}).

To obtain the $q$-parameterized Binet-Cauchy relation we put $\textbf{y}= \textbf{q}\equiv(q,q^2,\ldots,q^N)$, $\textbf{x}=\textbf{q}/q\equiv(1,q,\ldots, q^{N-1})$ in (\ref{schr}) and obtain:
 \begin{equation}
\sum_{\bla \subseteq
{M^N}}S_{\bla} (\textbf{q}) S_{\bla}
(\textbf{q}/q)=
\CV^{-1}_N(\textbf{q})
\CV^{-1}_N(\textbf{q}/q)\det \begin{pmatrix} \displaystyle{ \frac{1-q^{(M+N)(j+k-1)}}{1-q^{j+k-1}}}
\end{pmatrix}_{1\leq j, k \leq N}.
  \label{qschr}
\end{equation}
The relation (\ref{qschr})
is used in calculation of the amplitudes of the low temperature asymtotics of the correlation functions in the limit when the total number of sites is large enough, $M\gg 1$, while the number of particles $N$ is moderate: $1\ll N\ll M$ \cite{b5}. The
connection of this equation with the enumeration of plane partitions in a $N\times N\times M$ box in the fames of Quantum Inverse Scattering Method \cite{f, KBI2} was established in \cite{bogol}. We shall denote the box of the size $L\times N\times P$ as the set of integer lattice points:
\[{\cal B}(L, N, P)=\bigl\{(i, j, k)\in\BN^3
\bigl|\,0\leq i\leq L,\,\, 0\leq j\leq N,\,\,
0\leq k \leq P\bigr\}\,.\]
In \cite{bogol} the determinant in right-hand side of Eq.~(\ref{qschr}) was expressed as the Kuperberg determinant \cite{kup}, what led to the answer:
\begin{equation}\label{kup}
   \CV^{-1}_N(\textbf{q})
\CV^{-1}_N(\textbf{q}/q)\det \begin{pmatrix} \displaystyle{ \frac{1-q^{(M+N)(j+k-1)}}{1-q^{j+k-1}}}
\end{pmatrix}_{1\leq j,k \leq N}=\prod_{k=1}^{N} \prod_{j=1}^N
\frac{1-q^{M+j+k-1}}{1-q^{j+k-1}}.
 \end{equation}
This formula is the MacMahon generating function for the boxed plane partitions \cite{bres}.

As it follows from \cite{prep}, the sum of the Schur functions in left-hand side of (\ref{qschr}) may be expressed through the \textit{$q$-binomial determinant}:
\begin{equation}\label{sfqbd}
\sum_{\bla \subseteq
{M^N}}S_{\bla} (\textbf{q}) S_{\bla}
(\textbf{q}/q)=q^{\frac {NM}2 (1-M)}
\,\det \left(\displaystyle{
\begin{bmatrix} 2N+i-1 \\                                     N+j-1\end{bmatrix}}
  \right)_{1\leq i, j\leq M}\,.
\end{equation}
The entries in (\ref{sfqbd}) are the $q$-{\it binomial coefficients}, \cite{kac}, defined as
\begin{equation}\label{qbc}
  \begin{bmatrix}R\\r\end{bmatrix}\, \equiv\,
\,\frac{[R]!}{[r]!\,[R-r]!}\,,
\end{equation}
where $[n]$ is the $q$-{\it
number} being $q$-analogue of a
positive integer $n\in\BZ^+$,
$$
[n]\,\equiv\,\frac{1-q^n}{1-q}\,,
$$
and the $q$-{\it factorial} $[n]!$ is:
$[n]!\,\equiv\,[1]\,[2]\,\dots\,[n]$,
$[0]!\,\equiv\,1$. The determinant in right-hand side of (\ref{sfqbd}) was directly calculated in \cite{prep}, and the obtained answer agrees with (\ref{qschr}), (\ref{kup}).

\section{Schur functions and the lattice paths}

In this Letter we shall give the combinatorial interpretation of Eq.~(\ref{qschr}). This equation appears in the integrable models of strongly correlated bosons, \cite{bogol}, and of free fermions, \cite{b5}. It is well-known that a combinatorial description of the Schur functions may be given in terms of \textit{semistandard Young tableaux}. A filling of the cells of the Young diagram of $\bla$ with positive integers $n\in \BN^+$ is called a \textit{semistandard tableau of shape} $\bla$ provided it is weakly increasing along rows and strictly
increasing along columns.
The weight $\textbf{x}^T$ of a tableau $T$ is defined as
\begin{equation}
\nonumber
   \textbf{x}^T \equiv \prod_{i, j} x_{T_{ij}}\,,
\end{equation}
where the product is over all entries $T_{ij}$ of the tableau $T$. An equivalent definition of the Schur function is given by
\begin{equation}
\label{eqsch}
  S_{\bla}(x_1, x_2, \dots, x_m) =\sum_T \textbf{x}^T\,,
\end{equation}
where $m\geq N$, and the sum is over all tableaux $T$ of shape $\bla$ with the entries being numbers from the set $\{1, 2, \ldots, m\}$.

There is a natural way of representing each semistandard tableau of shape $\bla$ with entries not exceeding $N$ as a nest of  self-avoiding lattice paths with
specified start and end points. Let $T_{ij}$ be an entry in $i^{\rm th}$ row and $j^{\rm th}$ column of the
semistandard tableau $T$. The $i^{\rm th}$ lattice path of the nest $\emph{C}$
(counted from the top of the nest)
encodes the $i^{\rm th}$ row of the tableau ($i=1, \ldots, N$). It goes from $C_i=(N-i+1, N-i)$ to ($1, \mu_i=\la_i+N-i$) (see Fig.~2).
It makes $\la_i$ steps to the north so that the step along the line $x_j$ corresponds to the occurrences of the letter $N-j+1$ in the $i^{\rm th}$ row of $T$. The power $l_j$ of $x_j$ in the weight of any particular nest of paths is the number of steps to north taken along the vertical line $x_j$. Thus, an equivalent representation of the Schur function takes the form:
\begin{equation}\label{schrepr}
  S_{\bla} (x_1,x_2,\ldots,x_N) = \sum_{\emph{C}}\prod_{j=1}^{N}
x_{j}^{l_j},
\end{equation}
where summation is over all admissible nests $\emph{C}$. This representation of the Schur functions is natural in the Quantum Inverse Scattering Method approach to the solution of the models.
\begin{figure}[h]
\center
\includegraphics {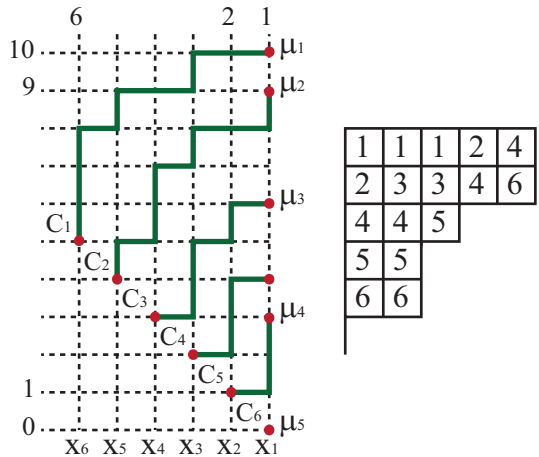}
\caption{A semistandard tableau of shape $\bla=(5,5,3,2,2,0)$ is represented  as a nest of lattice paths. Vertical steps along the line $x_j$ represent occurrences of letter $N-j+1$, $N=6$, in the tableau.}
\end{figure}
The $k^{\rm th}$ path is contained in a rectangle of the size $\la_k\times (N-k)$, $k=1,\ldots, N$. The starting point of each path is the lower left vertex. We define the volume of the path as the number of squares below it in the corresponding rectangle. The volume of the nest of lattice paths is equal to the volume of the lattice paths:
\begin{equation}
\nonumber
  \mid \zeta \mid_{\emph{C}}\,\,= \sum_{j=1}^{N}(j-1)l_j.
\end{equation}
Therefore, the $q$-parametrized Schur function is a partition function of the described nest:
\begin{equation}
\nonumber
  S_{\bla}(\textbf{q}/q) = \sum_{\emph{C}} q^{\mid \zeta \mid_{C}}\,,
\end{equation}
where summation is over all admissible nests $\emph{C}$. Adding the weight of partition $|\bla|=\sum_{k=1}^N \la_k$ to the volume of the nest, we obtain that
\begin{equation}
\nonumber
  \mid \xi\mid_{\emph{C}}\,=\mid \la \mid+\mid \zeta \mid_{\emph{C}}\,= \sum_{j=1}^{N}j\,l_j\,,
\end{equation}
and
\begin{equation}
\nonumber
  S_{\bla}(\textbf{q})= \sum_{\emph{C}} q^{\mid \xi\mid_{C}}=
   q^{\mid \la \mid}\sum_{\emph{C}} q^{\mid \zeta \mid_{C}}=
   q^{\mid \la \mid} S_{\bla}(\textbf{q}/q)\,.
\end{equation}

Consider a \textit{conjugated} nest of self-avoiding lattice paths (see Fig.~3) from $(1,\mu_i=\la_i+N-i)$ to  $B_i=(i,N+M-i)$. The $i^{\rm th}$ path consists of $M-\la_i$ steps to the north. The representation of the Schur function corresponding to the described nest is:
\begin{equation}
\label{rsf}
S_{\bla} (x_1, x_2, \ldots, x_N) = \sum_{\emph{B}}\prod_{j=1}^{N}
x_{j}^{M-l_j},
\end{equation}
where summation is over all admissible nests $\emph{B}$ of $N$ self-avoiding lattice paths.
\begin{figure}[h]
\center
\includegraphics {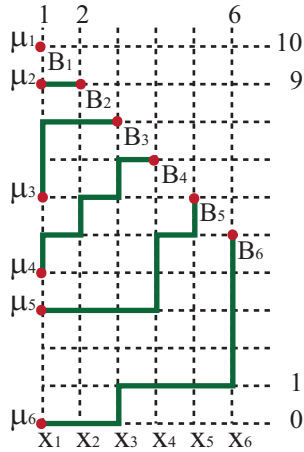}
\caption{Conjugated nest of lattice paths.}
\end{figure}
The $k^{\rm th}$ path is contained in a rectangle of the size $(k-1)\times M$, $k=1,\ldots,N$. The ending point of each path is the top right vertex. The volume of the path is the number of squares below it in the corresponding rectangle. The volume of the nest of the lattice paths is equal to the volume of the paths:
\begin{equation}
\nonumber
  \mid \zeta \mid_{\emph{B}}\,\,=\sum_{j=1}^{N} (j-1)(M-l_j).
\end{equation}

In the limit $q\rightarrow 1$, the Schur function is equal to the number of nests of self-avoiding lattice paths of the types either $\emph{B}$ or $\emph{C}$:
\begin{equation}
\nonumber
S_{\bla}(1, \ldots, 1) = \sum_{\emph{B}} 1 =\sum_{\emph{C}} 1\,.
\end{equation}
The summand of the scalar product (\ref{schr}), being the product of two Schur functions, may be graphically expressed as a nest of $N$ self-avoiding lattice paths starting at the equidistant points $C_i$ and terminating at the equidistant points $B_i$ ($i=1, \ldots, N$). This configuration, known as \textit{watermelon}, is presented on Fig.~4. The scalar product (\ref{schr}) is the sum of all such watermelons.
\begin{figure}[h]
\center
\includegraphics {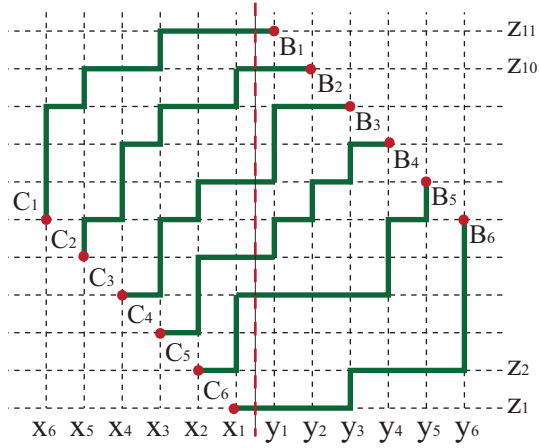}
\caption{Watermelon configuration}
\end{figure}
Repeating the arguments used above to derive the lattice paths volumes, it is straightforward to find that the volume of the watermelon is equal to:
\begin{equation}
\nonumber
  \mid w \mid=\mid\xi \mid_{\emph{C}}+ \mid\zeta\mid_{\emph{B}}\,.
\end{equation}
The partition function of watermelons (the generating function of watermelons) is equal to left-hand side of (\ref{qschr}):
\begin{equation}
\label{gfwm}
  {\sf W}(N,M)=\sum_{W}q^{\mid w \mid}=\sum_{\bla \subseteq
{M^N}}S_{\bla} (\textbf{q}) S_{\bla}
(\textbf{q}/q)\,,
\end{equation}
where the sum $\sum_{W}$ is taken over all watermelons with the fixed endpoints $C_i$, $B_i$, $1\le i\le N$.

To connect watermelon with a semistandard tableaux, let us now read the watermelon configuration with the endpoints  $C_i=(N-i+1,N-i)$, $B_i=(i,N+M-i)$ in the following way.
The $i^{\rm th}$ path (counted from the bottom) makes $\la_i=N$ steps to the east. The power $m_j$ of $z_j$ in the weight is the number of steps to the east taken along the horizontal line $z_j$. The Young tableau of such configuration is rectangle of the size $N\times N$. The Schur function of the watermelon is:
\begin{equation}\label{schwat}
  S_{\textbf{N}}(z_1, z_2, \ldots , z_{N+M}) = \sum_W\prod_{j=1}^{N+M} z_j^{m_j}\,,
\end{equation}
where summation is over all admissible watermelons, and ${\bf N}$ is the partition $(N, N, \ldots, N)$ of the length $N$, i.e., ${\bf N}\equiv N^N$ in our notations. The volume of watermelon is equal to
\begin{equation}
\nonumber
   \mid w \mid = \sum_{j=1}^{M+N}(j-1) m_j - \frac{N^2(N-1)}{2}.
\end{equation}
The partition function of watermelons is expressed through the Schur function (\ref{schwat}):
\begin{equation}\label{pfwm}
  {\sf W}(N,M) = q^{-\frac{N^2}2(N-1)}\, S_{\textbf{N}}(1, q^2, \ldots , q^{N+M-1})\,.
\end{equation}
This function is easy to calculate with the help of well known formula (see \cite{macd}, Chapter 1, Example 1):
\begin{equation}\label{schqpar}
  S_{\bla}(1,q^2,\ldots, q^{m-1}) = q^{n(\la)}\prod_{1\leq i<j\leq m} \frac{1-q^{\la_i-\la_j-i+j}}{1-q^{j-i}},
\end{equation}
where $n(\la)=\sum_i (i-1)\la_i$. Moreover, if $m>N$, then $\la_i=0$ for $i>N$. We obtain from (\ref{schqpar}) that
\begin{equation}\label{pfwmans}
  {\sf W}(N,M) = \prod_{i=1}^N \prod_{j=N+1}^{N+M} \frac{1-q^{N-i+j}}{1-q^{j-i}}\,.
\end{equation}
Replacing the indices $j\rightarrow N+j$ and $i\rightarrow N+1-i$, we put (\ref{pfwmans}) into the form:
\begin{equation}
\label{pfwmf}
   {\sf W}(N,M) = \prod_{i=1}^N\prod_{j=1}^{M} \frac{1-q^{N+i+j-1}}{1-q^{j+i-1}}=
  \prod_{i=1}^N \prod_{j=1}^{N} \frac{1-q^{M+i+j-1}}{1-q^{j+i-1}} \,.
\end{equation}
Eventually, it is seen that Eqs.~(\ref{gfwm}) and (\ref{pfwmf}) are in  agreement with Eqs.~(\ref{qschr}) and (\ref{kup}).

The watermelon with \textit{deviation} $k$ may be obtained by imposing the boundary condition $l_N=\ldots  =l_{N-k+1}=0$ in (\ref{schrepr}). The starting points $D_i$ of the watermelon with deviation will be shifted to the east by $k$ steps with respect to $C_i$. The watermelon with deviation is presented on Fig.~5. The boundary condition introduced is equivalent to the following property of the Schur function.
\begin{figure}[h]
\center
\includegraphics{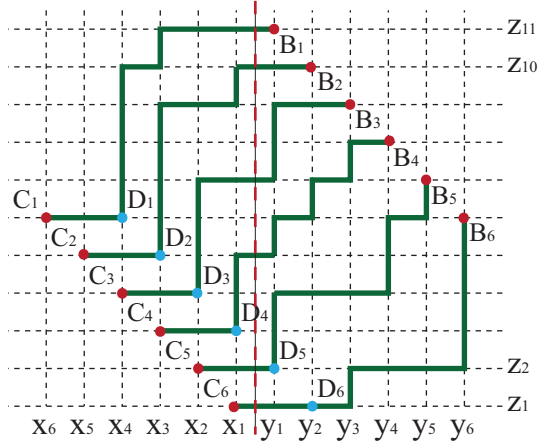}
\caption{Watermelon with deviation $k=2$. Starting points are $D_i$, endpoints are $B_i$. }
\end{figure}
Consider a partition $\bla=(\la_1,\ldots , \la_{N-k}, \la_{N-k+1}, \ldots, \la_{N})$ with the last $k$ parts equal to zero, $\la_{N-k+1}=\ldots =\la_{N}=0$. Then the limiting relation is valid:
\begin{equation}\label{limsch}
\lim_{x_N\rightarrow 0} \cdots \lim_{x_{N-k+1} \rightarrow 0} S_{\bla}(x_1, \ldots, x_{N-k}, x_{N-k+1}, \ldots, x_N) =
  S_{\tilde{\bla}}(x_1, \ldots, x_{N-k})\,,
\end{equation}
where the parts of $\tilde{\bla}=(\la_1, \la_2, \ldots , \la_{N-k})$ satisfy
$M \geq \la_1 \geq \la_2 \ldots \la_{N-k}\geq 0$. Taking the limit (\ref{limsch}) in (\ref{schr}), we obtain:
\begin{equation}
\sum_{\tilde\bla \subseteq
M^{N-k}}S_{\tilde{\bla}} (x_1, \ldots, x_{N-k}) S_{\hat{\bla}}
(y_1, \ldots, y_N)\,=\,\begin{pmatrix} \displaystyle{ \prod_{l=1}^{N-k}x_l^{-k}}\end{pmatrix}
\frac{\det (\tilde{M}_{k j})_{1\le
k, j\le N}}{\CV_{N-k}(\textbf{x})
\CV_N(\textbf{y})} \,,
\nonumber
\end{equation}
where summation is over all partitions $\tilde\bla$ with at most $N-k$ parts, each of which is less than or equal to $M$. The partition $\hat{\bla}$ of the length $N$ contains extra zeros $\hat{\lambda}_{N-k+1} = \hat{\lambda}_{N-k+2} = \ldots \hat{\lambda}_N=0$, and the entries $\tilde M_{kj}$ are:
\begin{equation}
\displaystyle{
\begin{array}{ll}
\tilde M_{kj} = M_{kj}\,, & 1\leq k \leq
N,\,\qquad 1\leq j \leq
N-k\,,\\[0.2cm] \tilde M_{kj} = y_j^{N-k}\,, & 1\leq k \leq N,\,\qquad\,N-k+1\leq j \leq N\,,
\end{array}} \nonumber
\end{equation}
where the entries $M_{kj}$ are given by (\ref{tt}).

The semistandard tableau corresponding to the watermelon with deviation consists of $N$ rows of the length $L=N-k$. The volume of the watermelon with deviation is
\begin{equation}
\label{wmdvol}
  \mid w \mid = \sum_{j=1}^{M+N}(j-1) m_j-\frac{N M (M-1)}{2}.
\end{equation}
In the case of the watermelon with deviation we obtain the representation analogous to (\ref{pfwm}):
\begin{gather}
\label{wmdpf}
{\sf W}(N,L,M) = q^{-\frac{NM(M-1)}{2}}\, S_{\textbf{L}}(1, q, \ldots, q^{N+M-1})\\[0.2cm]
  =\sum_{\tilde\bla \subseteq
M^{N-k}} S_{\tilde{\bla}} (q, \ldots, q^{N-k}) S_{\hat{\bla}}
(1, \ldots, q^{N-1})\,,
\nonumber
\end{gather}
where $\textbf{L}=L^N$ for the partition $\textbf{L}$. Calculating the Schur function $S_{\textbf{L}}$ with the help of (\ref{schqpar}), we obtain:
\begin{equation}\label{wmdans}
  {\sf W}(N,L,M) = \prod_{i=1}^N \prod_{j=N+1}^{N+M} \frac{1-q^{L-i+j}}{1-q^{j-i}}=
  \prod_{i=1}^N \prod_{j=1}^{M} \frac{1-q^{L+i+j-1}}{1-q^{j+i-1}}\,.
\end{equation}
In the limit $q\rightarrow 0$, this formula gives the number of the watermelons with deviation:
\begin{equation}\label{nwmde}
  A(N,L,M)=\prod_{i=1}^N \prod_{j=1}^{M}\frac{L+i+j-1}{j+i-1}\,.
\end{equation}

The Schur function can be expressed in a polynomial form through the complete symmetric functions, \cite{macd}:
$S_{\bla} ({\bf x})=\det (h_{\la_i -i+j}({\bf x}))_{1\leq i, j\leq N}$.
Under the $q$-parametrization, the complete symmetric functions are the $q$-binomial coefficients (\ref{qbc}):
\begin{equation}
\label{csf}
\displaystyle{h_r ({\bf q}/q)\,=\,\begin{bmatrix} N+r-1\\                                     r\end{bmatrix}}\,, \qquad 1\le r\le N.
\end{equation}

The following determinant with the $q$-binomial entries was calculated in \cite{igxv}:
\begin{equation}\label{qdetgv}
\det \left(q^{(j-1)(\la_i+j-i)}\displaystyle{
\begin{bmatrix} \la_i+m-i \\                                     m-j\end{bmatrix}}
  \right)_{1\leq i, j\leq N}=S_{\bla}(1,q,\ldots,q^{m-1})\,,\qquad m\ge N\,.
\end{equation}
Using (\ref{csf}) and the Pascal formula for the $q$-binomial coefficients,
\begin{equation} \label{pasc}
\begin{bmatrix}R\\r\end{bmatrix}\,=\,
\begin{bmatrix}R-1\\r-1\end{bmatrix}\,+\,
q^r\,
\begin{bmatrix}R-1\\r\end{bmatrix}\,,
\end{equation}
one can re-express left-hand side of (\ref{qdetgv}) so that the following equation holds:
\begin{equation}\label{qdetgv3}
\det \left(h_{\la_i-i+j}(1,q,\ldots,q^{m-1})
  \right)_{1\leq i, j\leq N}=S_{\bla}(1,q,\ldots,q^{m-1})\,.
\end{equation}

The partition function of the watermelon with deviation given by (\ref{wmdpf}) and (\ref{wmdans}) may be rewritten with regard to the determinantal formulas (\ref{qdetgv}) and (\ref{qdetgv3}):
\begin{eqnarray}
{\sf W}(N,L,M)&=&q^{-\frac{NM(M-1)}{2}}\det \left(q^{(j-1)(L+j-i)}\begin{bmatrix}
 L+M+N-i \\ M+N-j\end{bmatrix}
  \right)_{1\leq i, j\leq N}\label{qdetgv1}\\
  &=& q^{-\frac{NM(M-1)}{2}}\det \left(h_{L+j-i}({\bf q}/q)\right)_{1\leq i, j\leq N}
  \label{qdetgv2}\,.
\end{eqnarray}
The number of the watermelons with deviation (\ref{nwmde}) is expressed by
\begin{eqnarray}
 A(N,L,M)&=&\det \left(\begin{pmatrix}                                      L+M+N-i \\                                         M+N-j                                       \end{pmatrix}                                       \right)_{1\leq i, j\leq N} \label{wmdbd1}\\ [0.3cm] &=&
 \det \left(\begin{pmatrix}                                      L+M+N+j-i-1 \\                                         L+j-i                                       \end{pmatrix}                                       \right)_{1\leq i, j\leq N}\label{wmdbd2}\,,
\end{eqnarray}
where the determinant (\ref{wmdbd1}) is the {\it binomial determinant}, \cite{ges}, while
the coincidence of (\ref{wmdbd1}) and (\ref{wmdbd2}) can independently be checked by means of
(\ref{pasc}) at $q=1$.

In the limit $q\rightarrow 1$, the Schur function (\ref{rsf}) may be expressed with the help of (\ref{qdetgv}):
\begin{equation}\label{detgv}
  \det \left(\begin{pmatrix}                           \lambda_i+N-i \\  N-j \\                                       \end{pmatrix}
  \right)_{1\leq i, j\leq N}=S_{\bla}(1,\ldots,1)\,=\, \sum_{\emph{B}}1\,=\, \sum_{\emph{C}}1\,.
\end{equation}
Equation (\ref{detgv}) expresses the statement of the Gessel-Viennot theorem, \cite{ges}, connecting the binomial determinant in left-hand side of (\ref{detgv}) with the number of nests of self-avoiding lattice paths of the types either $\emph{B}$ or $\emph{C}$.

\section{Plane partitions and watermelons}

There exists bijection  between watermelons and plane partitions confined in a box of finite size \cite{4}.
A plane partition is an array $(\pi _{i j})_{1\leq i, j}$ of
non-negative integers that are
non-increasing as functions both of $i$ and $j$ \cite{macd, bres}. The integers $\pi_{i j}$ are called the parts of the plane partition, and $|\bpi| =\sum_{i, j} \pi_{i j}$ is its volume. Each plane partition has a three-dimensional diagram which can be
interpreted as a stack of unit cubes
(three-dimensional Young diagram). The
height of stack with coordinates $(i,j)$
is equal to $\pi_{i j}$. It is said that the plane partition corresponds to a box ${\cal B} (N, L, M)$ provided that
$j\leq N$, $i\leq L$ and $\pi_{i j} \leq M$ for all cubes of the Young diagram.
The generating function of plane partitions
\begin{equation}\label{gpp}
  Z_q (N,L,M)=\sum_{{\cal B} (N, L, M) } q^{\mid \bpi \mid}\,,
\end{equation}
where the sum is taken over all plane partitions contained in a box ${\cal B} (N, L, M)$.

Projection of gradient lines of plane partition (see Fig. 6) form a nest of lattice paths that correspond to watermelons (see Fig.~4 and Fig.~5, respectively).
\begin{figure}[h]
\center
\includegraphics {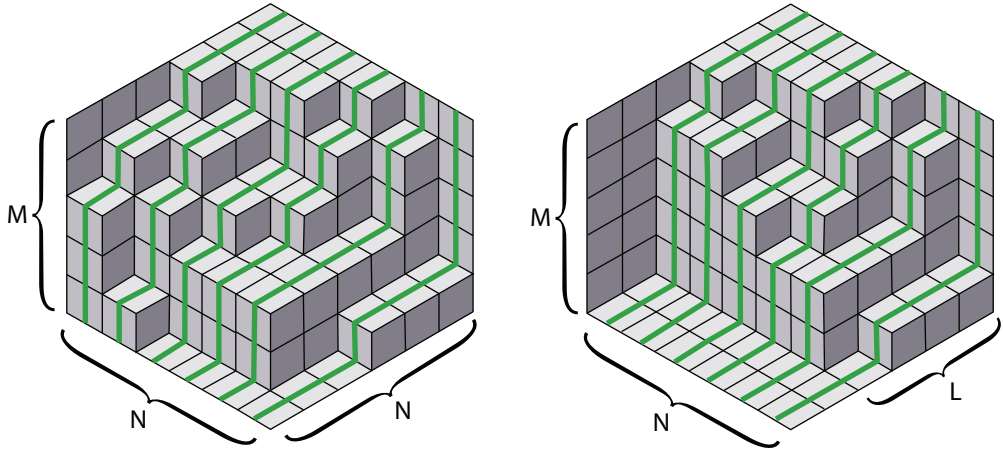}
\caption{Plane partitions with gradient lines embedded into a symmetric box ${\cal B} (N, N, M)$ and into an arbitrary one ${\cal B} (N, L, M)$, obtained as a special limit of symmetric box.}
\end{figure}
By its construction, the volume of watermelon (\ref{wmdvol}) coincides with the volume of plane partition $|\bpi|$, and thus
\begin{equation}
\nonumber
  Z_q(N,L,M)={\sf W}(N,L,M)\,.
\end{equation}

{\small \section* {Acknowledgement} Partially supported by RFBR (No.~13-01-00336).}


{\small\begin{thebibliography}{99}
\bibitem{macd}
            I.~G.~Macdonald,
            {\it Symmetric Functions and Hall Polynomials}, Oxford University Press, Oxford, 1995.
\bibitem{ful} W. Fulton, {\it Young Tableaux with Application to Representation Theory and Geometry}, Cambridge University Press, Cambridge, 1997.
\bibitem{bres}
           D.~M.~Bressoud, {\it Proofs and
           Confirmations. The Story of the
           Alternating Sign Matrix Conjecture},
           Cambridge University Press,
           Cambridge, 1999.
\bibitem{forr1}
           G.~Schehr, S.~N.~Majumdar, A.~Comtet, P.~J.~Forrester,
           {\it Reunion probability of N vicious walkers: typical and large fluctuations for large N},
            J. Stat. Phys. {\bf 149} (2012) 385-410.
\bibitem{zinn} P. Zinn-Justin, {\it Six-vertex model with domain wall boundary conditions and one-matrix model}, Phys. Rev. E {\bf 62} (2000) 3411-3418.
\bibitem{ok} A.~Okounkov, {\it Symmetric functions and random partitions},
In: Symmetric Functions 2001: Surveys of Developments and Perspectives,
NATO Science Series, Vol. {\bf 74} (2002) pp. 223-252.
\bibitem{hik} K.~Hikami, T. Imamura, {\it Vicious walkers and hook Young tableaux}, J. Phys. A: Math. Gen. {\bf 36} (2003) 3033–3048.
\bibitem{resh} A.~Okounkov, N.~Reshetikhin, {\it Correlation function of Schur process with application to local geometry of a random 3-dimensional Young diagram}, J. Amer. Math. Soc. {\bf 16} (2003) 581-603.
\bibitem{forr} G. T\'ellez, P.~J.~Forrester, {\it Expanded Vandermonde powers and sum rules
for the two-dimensional one-component plasma}, J. Stat. Phys. {\bf 148} (2012) 824–855.
\bibitem{bogol} N. M. Bogoliubov, {\it Boxed plane partitions as an exactly solvable boson model}, J. Phys. A: Math. Gen. \textbf{38} (2005) 9415-9430.
\bibitem{bogoltim}
N. M. Bogoliubov, J. Timonen, {\it Correlation functions for a strongly coupled boson system and plane partitions}, Phil. Trans. Roy. Soc. A \textbf{369} (2011) 1319-1333.
\bibitem{b1}
          N. M. Bogoliubov, {\it XX0 Heisenberg chain and random walks}, J. Math. Sci.
        \textbf{138} (2006) 5636-5643.
\bibitem{b2}
         N. M. Bogoliubov, {\it The integrable models for the vicious and friendly walkers}, J. Math. Sci. \textbf{143} (2007) 2729-2737.
\bibitem{b4}
           N.~M.~Bogoliubov, C.~Malyshev, {\it The correlation functions of the $XX$ Heisenberg magnet and random walks of vicious walkers},
           Theor. Math. Phys. \textbf{159} (2009) 563-574.
\bibitem{b5}
           N.~M.~Bogoliubov, C.~Malyshev, {\it The correlation functions of the $XXZ$ Heisenberg chain in the case of zero or infinite anisotropy, and random walks of vicioius walkers}, St.~Petersburg Math. J. \textbf{22} (2011) 359-377.
\bibitem{b6}
           N. M. Bogoliubov, C. Malyshev,
           \textit{Ising limit of a Heisenberg XXZ magnet and some temperature correlation functions}, Theor. Math. Phys. \textbf{169} (2011)         1517-1529.
\bibitem{prep}
           N.~M.~Bogoliubov, C.~Malyshev, {\it Correlation functions
            of XX0 Heisenberg chain, q-binomial determinants, and random walks},
            Nucl. Phys. B \textbf{879} (2014) 268-291.
\bibitem{f}
         L. D. Faddeev, \textit{Quantum completely integrable models of field theory}, Sov. Sci. Rev. Math. C, {\bf 1} (1980), 107–160; In: 40 Years in Mathematical Physics, World Sci. Ser. 20th Century Math., vol. 2, World Sci., Singapore, 1995, pp. 187–235.
\bibitem{KBI2}
            V.~E.~Korepin, N.~M.~Bogoliubov, A.~G.~Izergin,
     \textit{Quantum Inverse Scattering Method and Correlation
     Functions}, Cambridge University Press, Cambridge, 1993.
\bibitem{kup}
           G.~Kuperberg, {\it Another proof of the alternating sign martrix conjecture}, Int. Math. Res. Notices {\bf 1996} (1996) 139-150.
\bibitem{kac}
          A.~Klimyk, K.~Schmudgen, \textit{Quantum Groups and their Representations}, Springer, Berlin, 1997.
\bibitem{igxv}
I. Gessel, X. G. Viennot, {\it Determinants, paths, and plane partitions}, preprint (1989) 36~pp.
\bibitem{ges}
           I. Gessel, G. Viennot, {\it Binomial determinants, paths,and hook length formulae}, Advances in Mathematics, \textbf{58} (1985) 300-321.
\bibitem{4}
           A. J. Guttmann, A. L. Owczarek,
           X. G. Viennot,
           {\it Vicious walkers and Young tableaux~I: without walls},
           J. Phys. A: Math. Gen. \textbf{31} (1998) 8123-8135.
\end{thebibliography}}
\end{document}